\documentclass[twocolumn,showpacs,preprintnumbers,amsmath,amssymb]{revtex4}
\usepackage{graphicx}
\begin{document}
%\preprint{APS/123-QED}

%
% title page
%
\title{Role of initial quantum correlation in transient linear response }  
\author{Chikako Uchiyama\(^{1}\) and  Masaki Aihara\(^{2}\)}
\affiliation{%
\(^{1}\)Faculty of Engineering, University of Yamanashi,
4-3-11, Takeda, Kofu, Yamanashi 400-8511, JAPAN\\
\(^{2}\)Graduate School of Materials Science,
Nara Institute of Science and Technology\\
8916-5, Takayama-cho, Ikoma, Nara 630-0101 JAPAN}%
%\author{Chikako Uchiyama}
%\affiliation{%
%Faculty of Engineering, University of Yamanashi,\\
%4-3-11, Takeda, Kofu, Yamanashi 400-8511, JAPAN
%}%
\date{August 20,2010}  
\def\ch{{\cal H}}
\def\cha{{\cal H}_{S}}
\def\chb{{\cal H}_{R}}
\def\chp{{\cal H}_{P}}
\def\chab{{\cal H}_{SR}}
\def\ee{|1 \rangle \langle 1|}
\def\gg{|0 \rangle \langle 0|}
\def\eg{|1 \rangle \langle 0|}
\def\ges{|0 \rangle \langle 1|}
\def\muv{{\vec \mu}}
\def\ev{{\vec E}}
\def\hbar{\mathchar'26\mkern -9muh}
\def\ih{\frac{i}{\hbar}}
\def\dh{\frac{1}{\hbar}}
\def\v1{\mathbf{1}}
\def\tit{{\tilde t}}
\def\omegac{{\tilde \omega}_{c}}
\def\tbeta{{\tilde \beta}}
\def\am{A_{(m)}(t)}
\def\amt{A_{(m)}({\kern 1pt}{\tilde t}{\kern 1pt})}
\def\mm{{\vec \mu}_{(m)}(t)}
\def\mmt{\mu_{(m)}({\kern 1pt}{\tilde t}{\kern 1pt})}
\def\thm{\phi_{(m)}(t)}
\def\thmt{\phi_{(m)}({\kern 1pt}{\tilde t}{\kern 1pt})}
\def\Xim{\Xi_{(m)}(t)}
\def\xim{\xi_{(m)}(t)}
\def\com{c_{(m)}(t)}

%%%%%%%%%%% abstract %%%%%%%%%%%
\begin{abstract}
The linear transient response of a two-level system coupled with an environmental system is studied under correlated and factorized initial conditions.  We find that  the transient response in these cases differs significantly from each other, especially for strong system-environment interaction at intermediate temperatures.  This means that it is necessary to pay attention to the initial conditions chosen when analyzing experiments on transient linear response, because the conventional factorized initial condition results in an incorrect response, in which the quantum correlation between the relevant system and the environmental system is disregarded.
\end{abstract}
%%%%%%%%%%%%%%%%%%%%%%%%%%%%%

\maketitle
%%%%%%%%%%% sec.1 %%%%%%%%%%%
The linear response to an external field has been intensively studied to provide information on the characteristics of condensed matter. Since the early work on nuclear magnetic resonance (NMR) by Wangsness and Bloch \cite{Wangsness}, we often use a factorized initial condition of \(\rho_{S} \otimes \rho_{R}\), where the statistical operator \(\rho_{S}\) for the relevant spin system, and  that \(\rho_{R}\) for the environmental system separately stay in each equilibrium state.  This means that we conventionally neglect the quantum correlation between the system and the environment at the initial time which is justified only for a sufficiently weak system-environment interaction.  

We should pay attention to the fact that such factorization is sensitively reflected mainly in the transient behavior of the linear response to an external field, because the system feels the above mentioned difference in the initial condition after a sudden switch-on of the external field.  However, its effects have been discussed from various viewpoints such as (1) the reduced dynamics of an open quantum system\cite{Hakim,Ford88,Breuer95,Karrlein,Ford01,Ford01-2,Royer,vanKampen,Pollak}, (2) the complete positivity of dynamical maps\cite{Pechukas,Alicki,Royer-2,Stelmachovic,Jordan}, and (3) the nonlinear optical response\cite{Tanimura}.  The effect of a correlated initial condition on the linear response has been investigated by focusing only on the stationary susceptibility\cite{uchiyama09}.  Because the development of experimental techniques enables us to study transient behavior in a short time region, it might be important to investigate how the factorization of the initial condition affects the transient response under a suddenly applied external field.  In this paper, we answer this question using an exactly solvable model, i.e., a two-level system that linearly interacts with an environmental system (bosonic thermal bath).  

 The Hamiltonian of the matter system consists of a relevant two-level system, a bosonic bath, and the linear interaction between them, which are given in the form 
\begin{equation}
\ch= \cha+\chb+\chab,
\label{eqn:1}
\end{equation} 
with
\begin{eqnarray}
\cha &=& E_{1} \ee +E_{0} \gg, \nonumber \\
\chb &=& \sum_{k} \hbar \omega_{k} b_{k}^{\dagger}b_{k}, \label{eqn:2} \\
\chab &=&  \sum_{k} \hbar g_{k} (b_{k}^{\dagger}+b_{k}) \ee, \nonumber 
\end{eqnarray} 
where \(E_{0 (1) }\) is the energy of the lower (upper) state of the relevant system, \(\omega_{k}\) is the frequency of the bosonic bath mode, \(b_{k}^{\dagger}\) and \(b_{k}\) are its creation and annihilation operators, and \(g_{k}\) is the coupling strength between the excited state and the boson mode.  We study the transient behavior of a two-level system abruptly excited by an external field.   The matter-field interaction Hamiltonian is given by
\begin{equation}
\chp(t) = -\frac{1}{2} \muv \cdot \ev \; \theta(t)\eg e^{-i \omega_{p} t}  -\frac{1}{2} \muv^{*} \cdot \ev \; \theta(t) \ges e^{i \omega_{p} t} ,
\label{eqn:3} 
\end{equation} 
where \(\muv\) is the transition dipole moment, \(\ev\) is the amplitude of the external field, \(\omega_{p}\) is the frequency of the external field, and \(\theta(t)\) is the step function.  It is convenient to use a canonical transformation in terms of \(S \equiv \exp[B \ee]\) with \(B \equiv \sum_{k} (g_{k} / \omega_{k}) (b_{k}-b_{k}^{\dagger}) \), because this allows us to eliminate the system-bath interaction from the matter Hamiltonian in the form
\begin{eqnarray}
\ch'&=&S^{\dagger} \ch S = \cha'+\chb, \label{eqn:4} \\
\cha'&=&  E_{1}' \ee + E_{0} \gg,    \label{eqn:5} 
\end{eqnarray} 
with \(E_{1}' \equiv E_{1} -\hbar \sum_{k} (g_{k}^2 /\omega_{k})\). The matter-field interaction is transformed as
\begin{eqnarray}
\chp'(t)&=& S^{\dagger} \chp(t) S \nonumber \\
&=& -\frac{1}{2} \muv \cdot \ev \; \theta(t) \eg e^{-i \omega_{p} t} e^{B^{\dagger}} + h. c..
%&=& -\frac{1}{2} \muv \cdot \ev \theta(t) \eg e^{-i \omega_{p} t} e^{B^{\dagger}} -\frac{1}{2} \muv^{*} \cdot \ev \theta(t) \ges e^{i \omega_{p} t} e^{B} .
\label{eqn:6} 
\end{eqnarray} 
Under the application of an external field,  the time evolution of the density operator of the total system, \(\rho(t)\) for \(t>0\), is given by
\begin{equation}
\rho(t) = S e^{-\ih \ch' t} U(t) \rho'(0) U^{\dagger} (t) e^{\ih \ch' t} S^{\dagger},
\label{eqn:7} 
\end{equation} 
where we define \(\rho'(0) \equiv S^{\dagger} \rho(0) S\) and
\begin{equation}
U(t) \equiv T_{+} \exp[-\ih \int_{0}^{t} {\hat \chp}'(t') dt'], 
\end{equation} 
 with \( {\hat \chp}'(t) \equiv e^{\ih \ch' t} \chp'(t) e^{-\ih \ch' t}\).  Now we consider the linear response by making an approximation as
\begin{equation}
U(t) \approx 1-\ih \int_{0}^{t} {\hat \chp}'(t') dt'.
\end{equation} 

The initial statistical operator in the equilibrium state is transformed as
\begin{equation}
\rho'(0)= \rho'_{1} \ee + \rho'_{0} \gg,
\label{eqn:8} 
\end{equation} 
which gives the transformed density operator in the form
\begin{eqnarray}
\rho'(t) &=& S^{\dagger} \rho(t) S \nonumber \\
&&\hspace{-1.5cm}\approx e^{\ih \ch' t} [ \rho'_{1} \ee + \rho'_{0} \gg \nonumber \\
&& \hspace{-1.5cm} + \ih \int_{0}^{t} dt' \{ e^{i \Delta \omega t'}  \frac{1}{2} \muv \cdot \ev \; (G(t') \rho'_{0}- \rho'_{1}G(t')) \eg\nonumber \\
&&\hspace{-1cm}   + e^{- i \Delta \omega t'}  \frac{1}{2} \muv^{*} \cdot \ev \; (G^{\dagger}(t') \rho'_{1}- \rho'_{0}G^{\dagger}(t')) \ges \} ] e^{-\ih \ch' t}.\nonumber \\
\label{eqn:9} 
\end{eqnarray} 
where we define \(G(t) \equiv e^{B^{\dagger}(t)}\) with \(B^{\dagger}(t) \equiv e^{\ih \ch' t} B^{\dagger} e^{-\ih \ch' t}\) and \(\Delta \omega \equiv ( E_{1}' - E_{0} ) / \hbar - \omega_{p} \). 
The induced dipole moment under the application of an external field,
\begin{equation}
\muv(t)={\rm Tr}[(\muv \eg+\muv^{*} \ges) \rho(t)],
\label{eqn:10} 
\end{equation}
is evaluated by using Eq.~(\ref{eqn:9}) for each initial condition with trace operation for total system \({\rm Tr}\) which consists of trace operation over system (environmental) variables \({\rm Tr}_{S(R)}\).

As an initial condition of the total system, we consider the following two cases: (1) the total system is in an equilibrium state and (2) the two-level system and the environmental system are in separate equilibrium states. 

{\it Case (1): Correlated initial condition }\\
The initial state \(\rho(0)= \frac{1}{Z} \exp[-\beta \ch] \equiv \rho_{{\rm tot}}\) is transformed as \(\rho'(0)= \frac{1}{Z} \exp[-\beta \cha'] \exp[-\beta \chb] \) with \(Z={\rm Tr}[\exp[-\beta \ch]] \), which means that 
\begin{equation}
\rho'_{1} = \frac{e^{-\beta E_{1}'} \rho_{R}}{Z_{S}'}, \;\;\;\;
\rho'_{0} = \frac{e^{-\beta E_{0}} \rho_{R}}{Z_{S}'}, 
\label{eqn:11} 
\end{equation} 
where we defined  \(Z_{S}'={\rm Tr_{S} }[\exp[-\beta \cha']] \) and 
\(\rho_{R} =\frac{1}{Z_{R}} \exp[-\beta \chb]\) with \(Z_{R}={\rm Tr_{R} } \exp[-\beta \chb] \).  

{\it Case (2): Factorized initial condition}\\
For the factorized initial state \(\rho(0)= \rho_{S} \otimes \rho_{R}\), with \(\rho_{S}=\frac{1}{Z_{S}} \exp[-\beta \cha] \) and \(Z_{S}={\rm Tr_{S} } \exp[-\beta \cha] \), the transformation gives
\begin{equation}
\rho'_{1} = \frac{e^{-\beta E_{1}} G(0) \rho_{R} G^{\dagger}(0)}{Z_{S}}=\frac{e^{-\beta E_{1}} \rho_{R}'}{Z_{S}}, \;\;\;\;
\rho'_{0} = \frac{e^{-\beta E_{0}} \rho_{R}}{Z_{S}} ,
\label{eqn:12} 
\end{equation} 
with \( \rho_{R}'=\frac{1}{Z_{R}'} \exp[-\ch_{R}'] \) and \(Z_{R}'={\rm Tr_{R}} \exp[-\ch_{R}']\), where we define 
\begin{equation}
\ch_{R}'=\sum_{k} \hbar \omega_{k} (b_{k}^{\dagger}- \frac{g_{k}}{\omega_{k}}) (b_{k}-\frac{g_{k}}{\omega_{k}}).
\end{equation} 
This means that for the factorized initial condition, the excited state of the two-level system effectively interacts with a displaced bosonic reservoir.
It is noted that the initial statistical operator for case (1) (Eq.~(\ref{eqn:11})) has a factorized form, and that for case (2) (Eq.~(\ref{eqn:12})) has a correlated form.   This is simply because of the canonical transformation, and should not be confused  with correlated and factorized conditions in the original representation.
It is also noted that we have another choice of the initial decoupled state as \(\rho(0)= {\rm Tr_{S} }[\rho_{{\rm tot}}] \otimes {\rm Tr_{R}} [\rho_{{\rm tot}}]\), which will be discussed subsequently. 

Defining \(\mu_{(m)}(t) \) as an induced dipole moment for case \((m)\) with \(m=1,2\) apart from a factor of \(2\vec \mu (\muv^{*} \cdot \ev) / \hbar\), we obtain ,
\begin{equation}
\mu_{(m)}(t)  =   |\am| \cos (\omega _p t - \thm ),
\;\;\; (m=1,\;2) 
\label{eqn:13} 
\end{equation}
where \(\am\) is defined by the following equations, 
\begin{eqnarray}
A_{(1)} (t) &=&   \frac{1}{Z'_S } \int_{0}^{t} dt' e^{i \Delta \omega (t-t')} \{e^{-\beta E_{1}'} \Psi_{1}(t-t') \nonumber \\
&& \hspace{2cm} -e^{-\beta E_{0}} \Psi_{1}^{*} (t-t') \} ,\label{eqn:16}  \\
A_{(2)} (t) &=&  \frac{1}{Z_S } \int_{0}^{t} dt' e^{i \Delta \omega (t-t')} \{ e^{-\beta E_{1}} \Psi_{2}(t,t') \nonumber \\
&& \hspace{2cm} -e^{-\beta E_{0}} \Psi_{1}^{*} (t-t') \}, \label{eqn:17}  
%\Xi_{(1)} (t) &=&  \int_{0}^{t} dt' e^{i \Delta \omega (t-t')} \{e^{-\beta E_{1}'} \Psi_{1}(t-t')-e^{-\beta E_{0}} \Psi_{1}^{*} (t-t') \} ,\label{eqn:16}  \\
%\Xi_{(2)} (t) &=& \int_{0}^{t} dt' e^{i \Delta \omega (t-t')} \{ e^{-\beta E_{1}} \Psi_{2}(t,t') -e^{-\beta E_{0}} \Psi_{1}^{*} (t-t') \}. \label{eqn:17}  
\end{eqnarray}
and \(\thm\) is its argument.
With a coupling spectral function as \(h(\omega) \equiv \sum_{k} g_{k}^2 \delta(\omega-\omega_{k})\), the renormalized energy is given by \(E_{1}' = E_{1} -\hbar \int_{0}^{\infty} d\omega h(\omega)/\omega\).  \(\Psi_{1} (t)\) and \(\Psi_{2}(t,t')\) in Eq.~(\ref{eqn:16}) and Eq.~(\ref{eqn:17}) are obtained as 
\begin{eqnarray}
\Psi_{1} (t)&=&\exp[-\xi (t)- i \int_{0}^{\infty} d\omega \frac{h(\omega)} {\omega^2} \sin( \omega t )],  
\label{eqn:18} 
%\Psi_{1} (t)&=&\exp[-\int_{0}^{\infty} d\omega \frac{h(\omega)} {\omega^2} \{ (1+2 n(\omega)) (1-\cos(\omega t))+ i \sin( \omega t ) \}],  \label{eqn:18} \\
\end{eqnarray}
with \(\xi (t)=\int_{0}^{\infty} d\omega (h(\omega) / \omega^2) (1+2 n(\omega)) (1-\cos(\omega t))\) and 
\begin{eqnarray}
\Psi_{2} (t,t') &=& \Psi_{1} (t-t') \times \nonumber \\
&&\exp[-\int_{0}^{\infty} d\omega \frac{h(\omega)} {\omega^2} 2 i (\sin( \omega t')-\sin( \omega t)) ]. \nonumber \\ \label{eqn:19}
%\Psi_{2} (t,t') &=& \Psi_{1} (t-t')\exp[-\int_{0}^{\infty} d\omega \frac{h(\omega)} {\omega^2} 2 i (\sin( \omega t')-\sin( \omega t)) ]. \label{eqn:19}
\end{eqnarray} 

As is seen from Eq.~(\ref{eqn:11}) \(\sim\) ~(\ref{eqn:19}), the difference between the two cases in the transformed representation results in (1) a shift in the transition frequency \(\omega_{0}\), and (2) the displacement of the boson operator, which appears in the difference between \(\Psi_{1} (t)\) and \(\Psi_{2}(t,t')\) in Eq.~(\ref{eqn:18}) and ~(\ref{eqn:19}).

In the following, we show the numerical evaluations of the transient response for the Ohmic coupling spectral density, \(h(\omega)=s \omega e^{-\omega / \omega_{c}}\).   In this case, the renormalized energy is given by \(E_{1}' = E_{1} -\hbar s \omega_{c}\), and we find that the imaginary part of the exponent of Eq.~(\ref{eqn:18}) and ~(\ref{eqn:19}) is integrated with a formula as \(\int_{0}^{\infty} d\omega (h(\omega)/\omega^2) \sin( \omega t) =s \arctan \left( {\omega_c \,t} \right)\).  This means that the difference between the time evolution of \(\Psi_{1} (t)\) and \(\Psi_{2}(t,t')\) is written as \(\exp[ -2 i s (\arctan \left( {\omega_c \,t'} \right) - \arctan \left( {\omega_c \,t} \right))]\).
%\begin{equation}
%\Psi_{1} (t) = \exp[ - \phi (t) - i s \arctan \left( {\omega_c \,t} \right)],
%\label{eqn:17} 
%\end{equation}
%with \(\phi (t)=\int_{0}^{\infty} d\omega \frac{h(\omega)} {\omega^2} (1+2 n(\omega)) (1-\cos(\omega t))\) and 
%\begin{equation}
%\Psi_{2} (t,t') = \Psi_{1} (t-t')\exp[- 2 i s ( \arctan \left( {\omega_c \,t'} \right) -\arctan ( {\omega_c \,t}) ) ].
%\label{eqn:18} 
%\end{equation}

\begin{figure}[h]
\begin{center}
\includegraphics[scale=0.45]{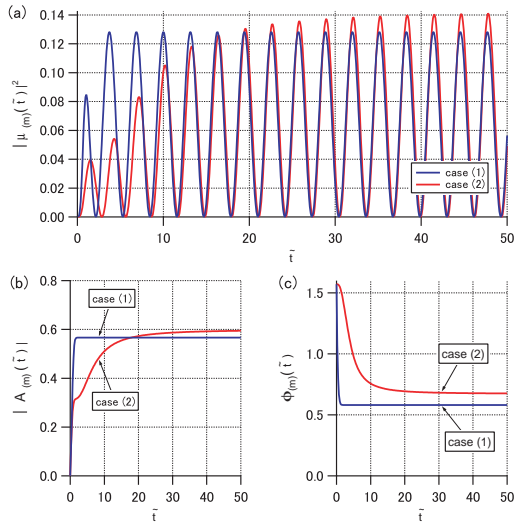}
\end{center}
\caption{(color online) Time evolution of the induced dipole moment for \(k_{B} T=10 \, \hbar \omega_{0}\), \(s=1\), \(\omega_{c} =\omega_{0}/5\), and \(\omega_{p}=\omega_{0}\). Dark gray (blue) and light gray (red) lines represent cases (1) and (2), respectively.  Time is scaled as \({\tilde t}=\omega_{0} t\). (a) time evolution of intensity of dipole moment \(|\mmt|^2\), (b) the time evolution of amplitude \(|\amt|\) and (c) the time evolution of phase, \(\thmt\) with \(m=1,2\).}
\label{fig:fig1}
\end{figure}

In Fig.~\ref{fig:fig1}, we show the time evolution of the intensity of the dipole moment under the application of an external field for \(k_{B} T=10 \,\hbar \omega_{0}\), \(s=1\), \(\omega_{c} =\omega_{0}/5\), and \(\omega_{p}=\omega_{0}\). Time is scaled as \({\tilde t}=\omega_{0} t\). 
%%%%%%%%%%%%%%% fig.1 %%%%%%%%%%%%%%%%%%%
%%%%%%%%%%%%%%% fig.1 %%%%%%%%%%%%%%%%%%%
We find that the evolution of case (1) approaches stationary oscillation much faster than that of case (2).  The characteristic time of the slower rise for case (2) depends on \(\omega_{c}\), which comes from the difference between \(\Psi_{1} (t-t')\) and \(\Psi_{2} (t,t')\), i.e., the \(\arctan \) part. We show in Fig.~\ref{fig:fig1}(b) the time evolution of amplitude, \(|\amt|\) with \(m=1,2\), where we find the amplitude for case (2) approaches larger value than that for case (1). The overestimation of the dipole moment for the factorized initial condition results from the fact that the decoherence due to quantum correlation is not taken into account in case (2). As shown in Fig.~\ref{fig:fig1}(c), we find that the two cases have different phases, \(\thmt\) with \(m=1,2\), for their oscillations. The feature originates from the effective displacement for the factorized initial condition shown in Eq.~(\ref{eqn:12}).  

\begin{figure}[h]
\begin{center}
\includegraphics[scale=0.45]{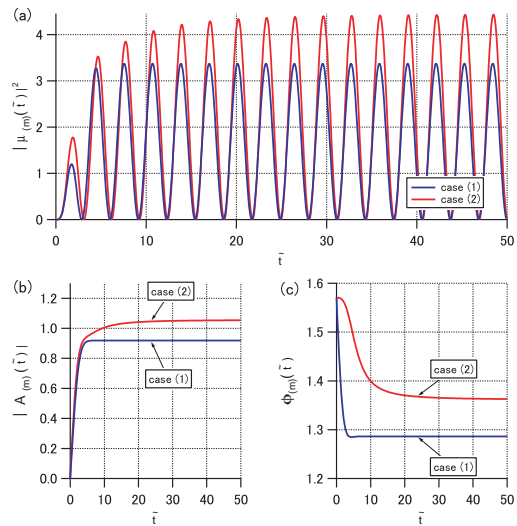}
\end{center}
\caption{(color online) Time evolution of the induced dipole moment for a lower temperature case \(k_{B} T=\hbar \omega_{0}\). Other parameters and evaluated quantities are the same as in Fig.~\ref{fig:fig1}.}
\label{fig:fig2}
\end{figure}

When the temperature is decreased to \(k_{B} T=\hbar \omega_{0}\), we find that the difference in the amplitudes of the stationary oscillations for the two cases increase (Fig.~\ref{fig:fig2}(a) and (b)).  And the asymptotic values of phase \(\thmt\) differ each other between \(m=1,2\).

%%%%%%%%%%%%%%% fig.2 %%%%%%%%%%%%%%%%%%%
%%%%%%%%%%%%%%% fig.2 %%%%%%%%%%%%%%%%%%%

\begin{figure}[h]
\begin{center}
\includegraphics[scale=0.45]{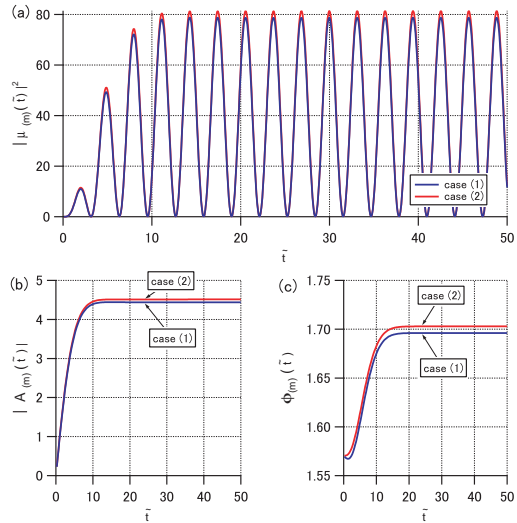}
\end{center}
\caption{(color online) Time evolution of the induced dipole moment for \(k_{B} T=\hbar \omega_{0}/5\). Other parameters and evaluated quantities are the same as in Fig.~\ref{fig:fig1}.}
\label{fig:fig3}
\end{figure}

However, as shown in Fig.~\ref{fig:fig3} for a lower temperature \(\hbar \omega_{0}/5\), we find that differences in amplitudes and phases of the two cases become very small. This is because at sufficiently low temperatures, the initial occupation probability for the upper level is very small, and the initial quantum correlation is disregarded.  
%%%%%%%%%%%%%%% fig.2 %%%%%%%%%%%%%%%%%%%
%%%%%%%%%%%%%%% fig.2 %%%%%%%%%%%%%%%%%%%

From these considerations, the effect arising from the initial correlation becomes significant in the intermediate temperature region, where the energy difference between the two levels becomes comparable to thermal excitation.

In order to see the difference in the transient response under the two types of initial conditions at the high temperature limit, we set \(n(\omega) \approx 1/ (\beta \hbar \omega)\), which gives an analytic form of  \(\xi (t)\) in Eq.~(\ref{eqn:18}) as 
\begin{equation}
\xi (t) \mathop  \to \limits^{\beta  \to 0} \frac{s}{2 \beta \hbar } (-4 \,t \,\arctan \left( {\omega_c \,t} \right) + \frac{{(2-\beta \hbar \omega_c) \log (1 + \omega_c ^2 \,t^2 )}}{\omega_c })
\end{equation}
 for the Ohmic spectral function.

%%%%%%%%%%%%%%% fig.3 %%%%%%%%%%%%%%%%%%%
\begin{figure}[h]
\begin{center}
\includegraphics[scale=0.45]{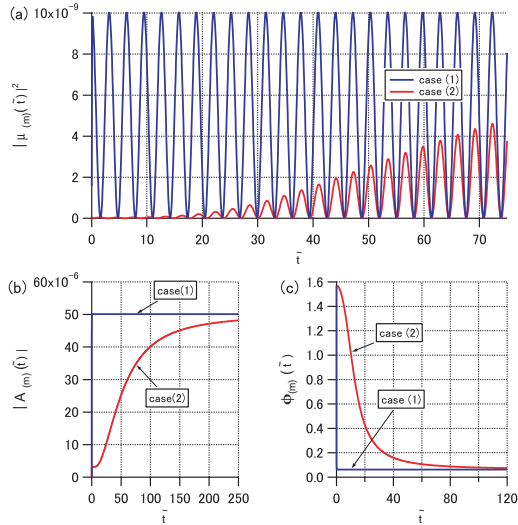}
\end{center}
\caption{(color online) Time evolution of the induced dipole moment for a high temperature case, \(k_{B} T=10^4 \hbar \omega_{0}\), with \( \omega_{c} =\omega_{0}/50\), \(s=1\), and \(\omega_{p}=\omega_{0}\).  The evaluated quantities are the same as in Fig.~\ref{fig:fig1}.}
\label{fig:fig4}
\end{figure}
%%%%%%%%%%%%%%% fig.2 %%%%%%%%%%%%%%%%%%%
In Fig.~\ref{fig:fig4} (a), we show the time evolution of intensity of dipole moment \(|\mmt|^2\) for \(k_{B} T=10^4 \hbar \omega_{0}\) with \( \omega_{c} =\omega_{0}/50\), \(s=1\), and \(\omega_{p}=\omega_{0}\). Fig.~\ref{fig:fig4} (b) and (c) show the time evolution of amplitude \(|\amt|\) and phase \(\thmt\) with \(m=1,2\), respectively for the same parameters as in Fig.~\ref{fig:fig4}(a). 

 We clearly find that a slow rise for case (2) for a smaller \(\omega_{c}\).  Generally,  \(1/\omega_{c}\) corresponds to the correlation time of the system-environment interaction, and the slow rise time determined by the long correlation time is an artifact arising from neglecting the initial quantum correlation.   
In contrast to the former cases of lower temperatures, we find that the asymptotic values of \(\amt\) and \(\thmt\) for \(m=1,2\) agree with each other very well in the long-time region.

In summary, we compared the transient linear response under two types of initial conditions, i.e., with and without initial correlation.  We found that the response for these cases differed from each other for strong system-environment interaction at intermediate temperatures.  Although the present analysis was limited to a two-level system linearly coupled with a bosonic environmental system, the initial quantum correlation may generally alter the linear responses of various types of materials.

\end{document}